\documentstyle[12pt]{article}
\begin{document}
\title{\bf Time-dependent backgrounds of two dimensional string theory from the $c=1$ matrix model}
\author{J. Sadeghi $^{a,}$\thanks{Email: pouriya@ipm.ir}\hspace{1mm}
and B.Pourhassan$^{a}$\thanks{Email: b.pourhassan@umz.ac.ir}\\
$^a$ {\small {\em  Sciences Faculty, Department of Physics, Mazandaran University,}}\\
{\small {\em P .O .Box 47415-416, Babolsar, Iran}}}
\maketitle
\begin{abstract}
\noindent The aim of this paper is to use correspondence between
solutions in the $c=1$ matrix model collective field theory and
coupled dilaton-gravity to a massless scalar field. First, we obtain
the incoming and outgoing fluctuations for the time-dependent
backgrounds with the lightlike and spacelike boundaries. In the case
of spacelike boundaries, we have done here for the first time. Then
by using the leg-pole transformations we find corresponding tachyon
field in two dimensional
string theory for lightlikes and spacelikes boundary.
\end{abstract}
\section{Introduction}
String theory in low space-time dimensions is an exactly solvable
theory. Indeed a theory with $1\leq D\leq2$ is a toy model for
solving many problems which haven't exact solutions in higher
dimensions. However, the  string theory in two dimension [1, 2, 3]
has more physics than other models. On the other hand, matrix model
[4, 5, 6] is a powerful mathematical tools to solve many problems of
two dimensional string theory. For example, by using the matrix
model one can obtain the simple linear equation of motion, while the
string theory yields to the non-linear equation. We would like to
consider the $2D$ string theory from the $c=1$ matrix model [7, 8, 9].\\
As we know there are $D-2$ transverse degrees of freedom for any
string. Therefore in two dimensional string theory there are no
transverse degrees of freedom, and only degree of freedom for the
on-shell field in this theory is the massless bosonic field $T(t,
\phi)$, which called the massless tachyon. We must note the graviton
and dilaton in two dimensional target space haven't the on-shell
degrees
of freedom and they are not physical particles.\\
Since the string coupling is an exponential function of space
coordinate $\phi$, then strings are free at the $\phi=-\infty$ and
there are no interactions between them, but they strongly coupled to
each other at $\phi=\infty$. On the other hand, existence of a
cosmological constant term, proportional to $e^{\phi}$, prevents the
creation of large positive terms of the $\phi$ in path integral. In
the other word there are a wall which strings (massless bosons or
tachyon fields) can't pass from it and scattered after the striking
to this wall. This wall often called Liouville or tachyonic wall.
One can consider such a system via the matrix quantum mechanics.
Indeed, the matrix model is a dual description of two dimensional
string theory. The matrix with large size is used in the matrix
model. For the $N\times N$ matrix there are $N$ pair of the $(x_{i},
y_{i})$ eigenvalues which distribute continuously on the two
dimensional surface at the $N\rightarrow\infty$ limit. This surface
give some information about particles density. Behavior of this
system is similar to the fermionic system, therefore one can
interpret the matrix model as a free fermionic system. Since the
$\frac{1}{N}$ factor is proportional to Plank's constant, $\hbar$,
then the $N\rightarrow\infty$ limit is corresponding to the
classical limit. In that case according to Liouville theorem,
particles in phase space move similar to an incompressible fluid
which known as the Fermi sea. This fermionic representation give the
microscopic description of two dimensional string theory, while the
macroscopic description is Das-Jevicki collective field theory [10].
Das-Jevicki collective field theory
includes all of the string interactions in two dimensions.\\
The matrix model description of two dimensional string theory is an
example of gauge/gravity correspondence, so the gauge theory lives
on the boundary of space-time in the gravity theory. Here we have
the matrix quantum mechanics as a gauge theory which at the
$N\rightarrow\infty$ limit lives in one dimension (time). On the
other hand we have Liouville string theory as a gravity theory which
lives in two dimensional target space-time. In Ref. [11] such a
correspondence considered, where exact solutions of the $c=1$ matrix
model collective field theory  and coupled dilaton-gravity to a
massless scalar field (tachyon field) have written and the relation
between scalar field at the lightlike boundary of space-time in both
theory have shown. Also they considered an simple example of the
time-dependent background with the lightlike boundary. In mentioed
Ref. Karczmarek and Wang have pointed the future work with the
spacelike boundaries. So, we take the motivation and progress from
their paper. Our goal in this paper is to consider other
time-dependent backgrounds, in particular the backgrounds with the
spacelike bounaries, and obtain  the incoming and outgoing
fluctuations. Then by using above correspondence and the leg-pole
transformations we will find tachyon fields. For this reason we will
consider several time-dependent backgrounds with the lightlike and
spacelike boundaries [12, 13, 14, 15, 16], and by using the same
method of Ref. [11] will obtain ratio of the outgoing to the incoming tachyon fields.\\
In the section 2 we review the correspondence between the $c=1$
matrix model and coupled dilaton-gravity to the massless scalar
field and also consider the leg-pole transformations which is
relation between the tachyon fields in the string theory and
fluctuations in the collective field theory [11]. Then in the
section 3 we consider solutions with the lightlike boundaries, so
after calculation of fluctuations we can obtain the tachyon fields.
In the section 4 we consider the solutions with the spacelike
boundaries and similar to the section 3 try to specify the tachyon
fields. Finally in the section 5 some conclusions and discussions
are given.
\section{ Relation Between The Matrix Model and $2D$ String Theory }
In the first part of this section, we have brief review to the $c=1$
matrix model and Liouville string theory of coupled dilaton-gravity
to the massless
scalar field [11], then in the second part, we describe relation between two theories.\\
As we know the $c=1$ matrix quantum mechanics describes a system of
free fermion with the harmonic oscillator potential. Eigenvalues of
the matrix form  discrete surfaces in two dimensional phase space
which called the Fermi sea. Boundary of the Fermi sea specified by
$p_{\pm}(x, t)$ which satisfy the condition $p_{+}(x_{l},
t)=p_{-}(x_{l}, t)$, where $x_{l}$ is the most left point of the
Fermi sea. Also $p_{\pm}$ are called the left and right chiral
components. Local density of fermions is given by difference between
$p_{+}$ and $p_{-}$. Static Fermi surface with the constant energy
$E=\mu$ have an equation as $(p-x)(p+x)=2\mu$, where $\mu$ must be
negative to have left and right branches without interaction. We
note here that any fluctuations along  the static background is
given by $\eta$ field. So fluctuations come from infinity to finite
$x$ and again come back to the infinity. In that case under
assumption $\mu >0$ we have [11],
\begin{equation}\label{s1}
p(x,t)=\sqrt{x^{2}-2\mu}+2\sqrt{\pi}\partial_{x}\eta.
\end{equation}
So for large negative $x$ and fluctuation field $\eta$ one can
write,
$x=\sqrt{2\mu}\cosh\sigma\approx\sqrt{\frac{\mu}{2}}e^{-\sigma}$ and
$\partial_{\sigma}\eta\approx|x|\partial_{x} \eta$, respectively.
Therefore, for given any time-dependent solution one can write the
corresponding solution in form of equation (1) to obtain $\eta$ as
an incoming fluctuation $\eta_{in}$. Then outgoing fluctuation given
by the following relation [11],
\begin{equation}\label{s2}
\eta_{out}=\eta_{in}-\frac{\sqrt{\pi}}{\mu}({\eta_{in}}^{\prime})^{2}
+\frac{2\pi}{3\mu^{2}}(\partial-1)({\eta_{in}}^{\prime})^{3}+{\mathcal{O}}((\eta_{in})^{4}).
\end{equation}
We are going to relate collective field theory $\eta$ to the tachyon
field in string theory. Indeed, by using results of Ref. [11], we
would like to obtain the tachyon fields for several lightlike and
spacelike backgrounds. In the coupled dilaton-gravity to a massless
scalar field there are three fields as dilaton, graviton and
tachyon. In this theory, we must note that, tachyon is the massless
and therefore isn't real tachyon.\\
At the zero-order there is linear dilaton background $\phi_{0}$, so
by definition $x^{\pm}=t\pm x$, we have $\phi_{0}=x^{+}-x^{-}=2x$.
Usually, such a background absorbs to the tachyon field $T$ by
definition of the new field $S=e^{-\phi_{0}}T$. One can expand
$S$ in higher order as $S=S^{(1)}+S^{(2)}+S^{(3)}+\ldots$.\\
Relations between the tachyon field $S$ in two dimensional string
theory and the fluctuation field $\eta$ in the collective field
theory are given by the leg-pole transformation [11],
\begin{eqnarray}\label{s3}
S_{in}(x^{-})&=&-\int{dv K(v-x^{-})\eta_{in}(v)},\nonumber\\
\eta_{in}(\sigma^{-})&=&-\int{dv K(\sigma^{-}-v)S_{in}(v)},\nonumber\\
S_{out}(x^{+})&=&\int{dv K(x^{+}-v+\ln{\frac{\mu}{2}})\eta_{out}(v)},\nonumber\\
\eta_{out}(\sigma^{-})&=&\int{dv
K(v-\sigma^{-}+\ln{\frac{\mu}{2}})S_{out}(v)},
\end{eqnarray}
where $x^{\pm}=t\pm x$ and $\sigma^{\pm}=t\pm \sigma$ are the
lightcone coordinates in the string theory and collective field
theory respectively. $K$ is a propagator which their integrals are
in terms of delta function [11]. The term of $\ln{\frac{\mu}{2}}$ in
relation (3) specifies position of Liouville wall. At the
$\mu\rightarrow0$ limit, Liouville wall is in depth of strong
coupling region, in that case one can neglect many scattering from
the tachyon background.\\
By using the relation (2) and putting $\eta_{out}$ order by order in
the leg-pole transformation (3) and find $S^{(1)}$, $S^{(2)}$ and
... separately, one can obtain the tachyonic field $S$. In two next
sections we will consider the lightlike and space like solutions,
and try to obtain the
tachyon field.\\
Before end of this section we would like to write General
time-dependent Fermi surface which have a following equation [14,
16],
\begin{equation}\label{s4}
x^{2}-p^{2}+\lambda_{-}e^{-rt}(x+p)^{r}+\lambda_{+}e^{rt}(x-p)^{r}+\lambda_{-}\lambda_{+}(x^{2}-p^{2})^{r-1}=2\mu,
\end{equation}
where $r$ is a non-negative integer, so $r=1,2$ is corresponding to
the classical solution of collective field theory and
$\lambda_{\pm}$ are finite constant parameters. We will consider
some special case of equation (4). Already the $r=1$ solutions in
Ref.s [11, 12, 17, 18, 19, 20] and $r=2$ solutions in Ref [14] are
discussed.
\section{Solutions With The Lightlike Boundaries}
In this section we consider three time-dependent solutions of the
Fermi surface and try to obtain corresponding tachyon field in two
dimensional string theory. We follow similar to Ref. [11] to obtain
the incoming and outgoing fluctuations $\eta$ and tachyon field $S$.
The simplest case of time-dependent background is given by [11],
\begin{equation}\label{s5}
(x+p+\lambda e^{t})(x-p)=2\mu,
\end{equation}
which is corresponding to the the equation (4) with $r=1$,
$\lambda_{-}=0$ and $\lambda_{+}=\lambda$. Here, $\lambda$ and $\mu$
are the positive constant. Equation (5) represents a moving
hyperbola which its center is at $(x,p)=(\lambda e^{t}, \lambda
e^{t})$. By choosing a parameter as $-\infty<\sigma<\infty$, we can
write $x=\sqrt{2\mu}\cosh\sigma-\frac{\lambda}{2}e^{t}$ and
$p=\sqrt{2\mu}\sinh\sigma-\frac{\lambda}{2}e^{t}$ as solutions of
the equation (5). For the large negative $\sigma$ we have $x-p\gg1$,
so one can write $x-p\approx2x$. Then for the large negative $x$ one
can rewrite the equation (5) as a following,
\begin{equation}\label{s6}
p\approx\sqrt{x^{2}-2\mu}-\lambda e^{t}.
\end{equation}
In the equation (6), we separate static and dynamics parts of the
equation (5), so for the large negative $x$ at $t=0$ we have
$x=\sqrt{\frac{\mu}{2}}e^{-\sigma}$ as expected. In order to obtain
the incoming fluctuation field we compare the equation (6) with the
equation (1) and find the following equation,
\begin{equation}\label{s7}
\eta_{in}\approx\frac{\lambda}{2}\sqrt{\frac{\mu}{2\pi}}e^{t-\sigma}.
\end{equation}
Then by using the equation (2) one can obtain outgoing fluctuation
field as,
\begin{equation}\label{s8}
\eta_{out}\approx\frac{1}{2\sqrt{\pi}}\left[\lambda
\sqrt{\frac{\mu}{2}}e^{t-\sigma}-\frac{\lambda^{2}}{4}e^2({t-\sigma})
+\frac{\lambda^{3}}{3}\sqrt{\frac{2}{\mu}}e^3({t-\sigma})\right].
\end{equation}
From the relation (7) it is clear that for the incoming fluctuations
we have $\eta_{in}\sim e^{t-\sigma}$, then by using the leg-pole
transformation (3) we can obtain the incoming tachyon field as
$S_{in}\sim e^{t-x}$. With the similar way one can obtain the
outgoing tachyon field for several order as $S_{out}^{(n)}\sim
(\frac{\mu}{2})^{n}e^{n(t+x)}$, with $n=1,2,3,...$. Therefor in the
first order, we can find the ratio of the outgoing to the incoming
tachyon fields as a following,
\begin{equation}\label{s9}
\frac{S_{out}}{S_{in}}\sim \mu e^{2x}=\mu e^{\phi_{0}}.
\end{equation}
It tell us that the ratio of the outgoing to the incoming tachyon
fields is
proportional to inverse of the string coupling.\\
Second case we consider in this section already introduced in [12,
17]. We use from the following equation of the Fermi surface and
will obtain the tachyon field. One can write the time-dependent
Fermi surface with the lightlike boundary which represents a moving
hyperbola as,
\begin{equation}\label{s10}
(x+p+2\lambda_{-}e^{-t}\frac{x+p}{x-p}+2\lambda_{+}e^{t})(x-p)=2\mu,
\end{equation}
where $\lambda_{\pm}$ is arbitrary non-negative constant. The center
of hyperbola (10) placed at $(x,p)=(-\lambda_{+} e^{t}-\lambda_{-}
e^{-t}, -\lambda_{+} e^{t}-\lambda_{-} e^{-t})$. Equation (10) is
corresponding to the general equation (4) with $r=1$ and
infinitesimal $\lambda_{\pm}$, so one can neglect the second order
of $\lambda_{\pm}$. Here, there are some special cases, for example
the solutions with $\lambda_{-}=0$ and
$\lambda_{+}=\frac{\lambda}{2}$, are the same as solutions of
equation (5).\\
We would like to consider other special case with
$\lambda_{+}=\lambda_{-}=\lambda$. In that case at the $t=0$ filled
part of Fermi sea is centered at $(x,p)=(-2\lambda,0)$. After the
time revolution, fermions (tachyon fields) coming from $x=-\infty$
to the finite point $x=-2\lambda-\sqrt{2\mu}$, then coming back to
$x=-\infty$. Similar to previous case and under assumption of very
small $\lambda$, one can rewrite the equation (10) as,
\begin{equation}\label{s11}
p\approx\sqrt{x^{2}-2\mu}-2\lambda
e^{t}-\frac{\mu\lambda}{x^{2}}e^{-t}.
\end{equation}
Then by comparing the equations (1) and (11), the incoming
fluctuation field will obtained as a following,
\begin{equation}\label{s12}
\eta_{in}\approx\lambda\sqrt{\frac{\mu}{2\pi}}(e^{t-\sigma}+e^{-(t+\sigma)}),
\end{equation}
and by using the equation (2) one can obtain the outgoing
fluctuation field as,
\begin{equation}\label{s13}
\eta_{out}\approx\frac{1}{\sqrt{\pi}}\left[\lambda
\sqrt{\frac{\mu}{2}}(e^{t-\sigma}+e^{-(t+\sigma)})-\frac{\lambda^{2}}{2}(e^{t-\sigma}
+e^{-(t+\sigma)})^{2}-\frac{\lambda^{3}}{3}\sqrt{\frac{2}{\mu}}(e^{t-\sigma}+e^{-(t+\sigma)})^{3}\right].
\end{equation}
Now to calculate tachyon field we use from the leg-pole
transformations (3). For incoming and outgoing tachyon field we find
that $S_{in}\sim e^{-x}(e^{t}+e^{-t})=2e^{-x}\cosh{t}$, and
$S_{out}^{(n)}\sim \mu e^{nx}(e^{t}+e^{-t})^{n}=2^{n}\mu
e^{nx}\cosh^{n}t$, respectively. Therefore in the first order we
find that the ratio of the outgoing to the incoming tachyon fields
is proportional to inverse of the string coupling. We must note that
our result for the tachyon field is agree with results of Ref. [12].
We must note that in the higher order of the outgoing tachyon field
the ratio of the outgoing to the incoming
tachyon fields is depend to the space-time coordinates.\\
Finally we consider the other interesting solution which satisfy the
equation of motion. It will be written as,
\begin{equation}\label{s14}
(x+p+2\lambda_{-}\frac{(x+p)^{2}}{x-p}+2\lambda_{+}e^{t})(x-p)=2\mu.
\end{equation}
Just like to previous case we assume that
$\lambda_{+}=\lambda_{-}=\lambda$ and then for large $x-p$ one can
obtain,
\begin{equation}\label{s15}
p\approx\sqrt{x^{2}-2\mu}-2\lambda e^{t}.
\end{equation}
So the incoming fluctuation obtained as the following relation,
\begin{equation}\label{s16}
\eta_{in}\approx\lambda\sqrt{\frac{\mu}{2\pi}}(e^{t-\sigma}).
\end{equation}
Again similar to previous case one can write the outgoing
fluctuation and then by using the leg-pole transformation (3) obtain
the corresponding tachyon fields. After all one can obtain the ratio
of outgoing to incoming tachyon fields proportional to inverse of
the string coupling.
\begin{equation}\label{s17}
\frac{S_{out}}{S_{in}}\sim \mu e^{\phi_{0}},
\end{equation}
which is similar to the equation (9), therefore solution (14) behave
as first case of this section given by the equation (5).\\
In the next section we will consider solutions with the spacelike
boundaries.
\section{Solutions With The Spacelike Boundaries}
Our main goal of this paper is to consider the Fermi surface with
the spacelike boundary [14, 15]. In this section we consider three
cases of such solutions and will try to obtain corresponding tachyon
field
in two dimensional string theory.\\
The first case is a time-dependent Fermi surface with the spacelike
boundary as a following [14],
\begin{equation}\label{s18}
(x+p-e^{2t}(x-p))(x-p)=2\mu,
\end{equation}
which is corresponding to the equation (4) with $r=2$,
$\lambda_{-}=0$ and $\lambda_{+}=-1$. The equation (18) represents a
closed hyperbola. If we set $\lambda=e^{t}(p-x)$ in the equations
(5) we will arrive to the equation (18). Also we can see now the
equivalence relation with the explicit calculation. For the large
negative $x$ one can rewrite the equation (18) as,
\begin{equation}\label{s19}
p\approx\sqrt{x^{2}-2\mu}+2|x| e^{2t}.
\end{equation}
As before, we compare it with the equation (1), so, for the incoming
fluctuation field we find the following expression,
\begin{equation}\label{s20}
\eta_{in}\approx\frac{\mu}{2\sqrt{\pi}}e^{2(t-\sigma)},
\end{equation}
and the outgoing fluctuation field easily obtained as,
\begin{equation}\label{s21}
\eta_{out}\approx\frac{\mu}{4\sqrt{\pi}}\left[e^{2(t-\sigma)}-e^{4(t-\sigma)}-3e^{6(t-\sigma)}\right].
\end{equation}
From the equation (20) we see the incoming fluctuation field
obtained as $\eta_{in}\sim e^{2(t-\sigma)}$, therefore by using the
relations (3) it is clear that $S_{in}\sim e^{2(t-x)}$. The outgoing
tachyon field for any order is obtained as $S_{out}^{(n)}\sim
(\frac{\mu}{2})^{2n}e^{2n(t+x)}$. In the first order we see that the
ratio of the outgoing to the incoming tachyon fields is proportional
to inverse of the squared string coupling.
\begin{equation}\label{s22}
\frac{S_{out}}{S_{in}}\sim \mu^{2}e^{2\phi_{0}}.
\end{equation}
Already, we use from $x\sim\cosh\sigma$, but it is valid at the
$x\rightarrow-\infty$ limit only. Generally, there are Alexandrov
coordinates transformations as,
\begin{eqnarray}\label{s23}
x&=&\sqrt{2\mu}\frac{\cosh{\sigma}}{\sqrt{1-e^{2\tau}}}\approx\sqrt{\frac{\mu}{2}}\frac{e^{-\sigma}}{\sqrt{1-e^{2\tau}}},\nonumber\\
t&=&\tau-\frac{1}{2}\ln(1-e^{2\tau}),
\end{eqnarray}
which at the $\tau\rightarrow-\infty$ limit reduce to
$x=\sqrt{2\mu}\cosh\sigma$ and $t\rightarrow-\infty$, as expected.
In terms of coordinates (23), the incoming fluctuation field
obtained as,
\begin{equation}\label{s24}
\eta_{in}\approx\frac{\mu}{4\sqrt{\pi}}\frac{e^{2(\tau-\sigma)}}{(1-e^{2\tau})^{2}}.
\end{equation}
The equation (24) reduces to the equation (20) at the
$\tau\rightarrow-\infty$ limit, where $t=\tau$. In that case to
obtain the outgoing fluctuation field $\eta_{out}$ from the relation
(24) we note that all derivative in the equation (2) are in terms of
$\sigma$, and $\sigma$-dependent term in (24) is as before [equation
(20)]. Therefore after calculation of the incoming and outgoing
tachyon fields we will have similar solution with the equation (22).\\
The second time-dependent fermi surface with the spacelike boundary
which represents an open hyperbola is given by [14],
\begin{equation}\label{s25}
(x+p+e^{2t}{(x-p)})(x-p)=2\mu,
\end{equation}
which is corresponding to $r=2$, $\lambda_{-}=0$ and $\lambda_{+}=1$
in general solution (4). Static solution of the equation (25) is
obtained at the $t\rightarrow-\infty$ limit. Here we will find the
same solution as previous case [see equations (20), (21) and (22)].
But in here, Alexsandrov coordinates is different with the first
case [14]. By using the equation (25) one can obtain the incoming
fluctuation field in terms of Alexsandrov coordinates as,
\begin{equation}\label{s26}
\eta_{in}\approx\frac{\mu}{4\sqrt{\pi}}\frac{e^{-2(\tau+\sigma)}}{(1-e^{-2\tau})^{2}},
\end{equation}
which at the $\tau\rightarrow-\infty$ limit, where $t=-\tau$,
reduces to the expected relation (20).\\
Now we consider the third case of the time-dependent Fermi surface
given by the following equation [12, 13],
\begin{equation}\label{s27}
e^{-2t}(x+p-\lambda_{+}e^{t})^{2}+e^{2t}(x-p-\lambda_{-}e^{-t})^{2}=2\mu.
\end{equation}
We are going to consider special case with
$\lambda_{+}=\lambda_{-}=-1$. In that case equation (27), for
infinitesimal $\mu$, represents a closed Fermi surface which at the
initial time $t=0$ is a circle with the radius $\sqrt{2\mu}$
centered in $(x,p)=(-1,0)$. After the time revolution this circle
changes to an moving ellipse. In here there are the condition
$\lambda> \sqrt{2\mu}$
[13], which under assumption of $\mu\ll1$ will satisfied.\\
To write the equation (27) in the form of the equation (1) we note
that the described universe by the equation (27) is in weak coupling
at both early and late times. It means that the situation with
$t\rightarrow-\infty$ and $x-p\gg1$ is equivalent to the situation
with $t\rightarrow0$ and $x+p\gg1$. Therefore, after rescaling
$2(\mu-1)\rightarrow\mu$ and for the large negative $x$ one can
rewrite the equation (27) as a following,
\begin{equation}\label{s28}
p\approx\sqrt{x^{2}-2\mu}+4 e^{t}(1-|x|e^{t}).
\end{equation}
As a result one can find the incoming fluctuation field as,
\begin{equation}\label{s29}
\eta_{in}\approx-\frac{1}{\sqrt{\pi}}\left[\sqrt{2\mu}e^{t-\sigma}+\frac{\mu}{\sqrt{2}}e^{2(t-\sigma)}\right],
\end{equation}
and the outgoing fluctuation field to second order obtained as,
\begin{equation}\label{s30}
\eta_{out}\approx-\frac{1}{\sqrt{\pi}}\left[\sqrt{2\mu}e^{t-\sigma}+\frac{\mu}{\sqrt{2}}e^{2(t-\sigma)}\right]
-\frac{1}{\mu\sqrt{\pi}}\left[\sqrt{2\mu}e^{t-\sigma}+\frac{\mu}{\sqrt{2}}e^{2(t-\sigma)}\right]^{2}.
\end{equation}
by using equations (29), (30) and the leg-pole transformation (3) we
will find the tachyon field in terms of both lightlike and spacelike
tachyon field which obtained already in the equations (5) and (18).
Indeed for the incoming tachyon field one can obtain $S_{in}\sim
e^{t-x}+e^{2(t-x)}$ and for outgoing tachyon field at the first
order one can obtain $S_{out}\sim \mu e^{t+x}+\mu^{2}e^{2(t+x)}$.
Here we see that, even at the first order, the ratio of the outgoing
to the incoming tachyon fields obtained in terms of space-time
coordinates.
\section{Conclusion}
In this paper we consider several time-dependent Fermi surface with
the null and spacelike boundaries and obtained the tachyon fields by
using correspondence between the $c=1$ matrix model and two
dimensional string theory. Already in Ref. [11] simple
time-dependent background with the lightlike boundary considered.
Here we calculated ratio of the outgoing to incoming tachyon fields
for other time dependent backgrounds with the lightlike and
spacelike boundaries. In the case of time-dependent Fermi sea with
null boundaries we found that ratio of the outgoing to the incoming
tachyon fields is proportional to inverse of the string coupling.
Same calculations for solutions with the spacelike boundaries
[equations (18) and (25)] show that the ratio of the outgoing to the
incoming tachyon fields is proportional to inverse of the squared
string coupling. Finally for a solution in the form of equation
(27), which defined closed cosmological universe, we obtained a
combining tachyon field proportional to the lightlike and spacelike
tachyon fields. In summary we considered three solutions of the
Fermi sea with lightlike boundaries in equations (5), (10) and (14),
and three solutions of the Fermi sea with spacelike boundaries in
equations (18), (25) and (27), then obtained the tachyon fields
corresponding to the fluctuation fields.\\
It may be interesting to consider any time-dependent solutions such
as $x^{2}-p^{2}=1+(x-p)^{3}e^{3t}$ [13], which is corresponding to
$r=3$ in relation (4), and try to obtain tachyon field.

\end{document}